\documentclass[journal,12pt,onecolumn,draftclsnofoot,a4paper,]{IEEEtran}
\usepackage[T1]{fontenc}
\usepackage[utf8]{inputenc}
\usepackage{amsmath}
\usepackage{amssymb}
\usepackage{graphicx}
\usepackage{acronym}
\usepackage{theorem}
\usepackage{multirow}
\usepackage{verbatim}
\usepackage{booktabs}
\usepackage{color}
\usepackage{subfigure}
\usepackage{cite}
\usepackage{tikz}
\usepackage{enumerate}
\usepackage{comment}
\usepackage{enumitem}
\usepackage[acronym,shortcuts]{glossaries}
\usepackage{mathtools}
\usepackage{cleveref}

\usepackage{algorithm}
\usepackage{algpseudocode}

\algdef{SE}[DOWHILE]{Do}{doWhile}{\algorithmicdo}[1]{\algorithmicwhile\ #1}%

\newcommand{\bH}{\mathbf{H}}
\newcommand{\bG}{\mathbf{G}}
\newcommand{\bS}{\mathbf{S}}

\newcommand{\be}{\mathbf{e}}
\newcommand{\ba}{\mathbf{a}}
\newcommand{\bb}{\mathbf{b}}
\newcommand{\bm}{\mathbf{m}}
\newcommand{\bs}{\mathbf{s}}
\newcommand{\bh}{\mathbf{h}}
\newcommand{\bx}{\mathbf{x}}

\newcommand{\bA}{\mathbf{A}}
\newcommand{\bB}{\mathbf{B}}

\newcommand{\bzero}{\mathbf{0}}

\newcommand{\modp}{\hspace{1mm}\mathrm{mod}\hspace{1mm}p}

\definecolor{mygreen}{rgb}{0.0, 0.72, 0.24}
\definecolor{myred}{rgb}{0.89, 0.0, 0.13}
\definecolor{myorange}{rgb}{0.93, 0.53, 0.18}

{\theorembodyfont{\rmfamily}
   \newtheorem{Def}{{\textbf Definition}}}
   
      \newtheorem{Lem}{Lemma}

      \newtheorem{Ass}{Assumption}
      \newtheorem{Pro}{Proposition}

   \definecolor{mybrown}{rgb}{1,0.5,0.3}
	
	\newacronym{QC}{QC}{quasi-cyclic}
\newacronym{DF}{DF}{difference family}
\newacronym{BF}{BF}{Bit Flipping}
\newacronym{DFR}{DFR}{Decoding Failure Rate}
\newacronym{LDPC}{LDPC}{Low-Density Parity-Check}
\newacronym{QC-LDPC}{QC-LDPC}{Quasi-Cyclic Low-Density Parity-Check}
\newacronym{QC-MDPC}{QC-MDPC}{Quasi-Cyclic Moderate-Density Parity-Check}
\newacronym{MDPC}{MDPC}{Moderate-Density Parity-Check}
\newacronym{CCA}{CCA}{Chosen Ciphertext Attack}
\newacronym{CPA}{CPA}{Chosen Plaintext Attack}
\newacronym{CPP}{CPP}{Cliques Partition Problem}
\newacronym{SDP}{SDP}{Syndrome Decoding Problem}
\newacronym{ISD}{ISD}{Information Set Decoding}
\begin{document}
\title{Analysis of reaction and timing attacks against cryptosystems based on sparse parity-check codes} 
%
%
\author{\IEEEauthorblockN{Paolo Santini, Massimo Battaglioni, Franco Chiaraluce and Marco Baldi}
\IEEEauthorblockA{Dipartimento di Ingegneria dell'Informazione\\
Universit\`a Politecnica delle Marche\\
Ancona, Italy\\
Email: \{p.santini\}@pm.univpm.it, \{m.battaglioni, f.chiaraluce, m.baldi\}@univpm.it}}
%
%
%
\maketitle              
\begin{abstract}
In this paper we study reaction and timing attacks against cryptosystems based on sparse parity-check codes, which encompass low-density parity-check (LDPC) codes and moderate-density parity-check (MDPC) codes.
We show that the feasibility of these attacks is not strictly associated to the quasi-cyclic (QC) structure of the code but is related to the intrinsically probabilistic decoding of any sparse parity-check code. So, these attacks not only work against QC codes, but can be generalized to broader classes of codes.
We provide a novel algorithm that, in the case of a QC code, allows recovering a larger amount of information than that retrievable through existing attacks and we use this algorithm to characterize new side-channel information leakages.
We devise a theoretical model for the decoder that describes and justifies our results. Numerical simulations are provided that confirm the effectiveness of our approach. 
\end{abstract}

\begin{IEEEkeywords}
Code-based Cryptosystems, LDPC codes, MDPC codes, Reaction Attacks, Timing Attacks
\end{IEEEkeywords}

\section{Introduction}
The code-based cryptosystems introduced by McEliece~\cite{McEliece1978} and Niederreiter~\cite{Niederreiter1986} are among the oldest and most studied post-quantum public-key cryptosystems.
They are commonly built upon a family of error correcting codes, for which an efficient decoding algorithm is known. 
The security of these systems is based on the hardness of the so called \ac{SDP}, i.e., the problem of decoding a random linear code, which has been proven to be NP-hard~\cite{Berlekamp1978}. 
The best \ac{SDP} solvers are known as \ac{ISD} algorithms and were introduced in 1962 by Prange~\cite{Prange1962}; improved through the years (see~\cite{Lee1988, Stern1989, Becker2012}, for some well known variants), these algorithms are characterized by exponential complexity, even when considering adversaries equipped with quantum computers~\cite{Bernstein2010}. 
Because of their well studied and assessed security, code-based cryptosystems are nowadays considered among the most promising candidates for the post-quantum world \cite{NISTreport2016}.

In the above schemes, and others of the same type, the private key is the representation of a code, whose parameters are chosen in such a way to guarantee decoding of a given amount of errors, which are intentionally introduced in the plaintext during encryption.
The public key is obtained through linear transformations of the secret key, with the aim of masking the structure of the secret code.
In the original McEliece proposal, Goppa codes were used: on the one hand, this choice leads to a well assessed security (the original proposal is still substantially unbroken); on the other hand, the corresponding public keys do not allow for any compact representation, and thus have very large sizes.

A well-known way to address the latter issue is that of adding some geometrical structure to the code, in order to guarantee that the public key admits a compact representation. 
The use of \ac{QC} codes with a sparse parity-check matrix naturally fits this framework: the sparsity of the parity-check matrix allows for efficient decoding techniques, while the quasi-cyclic structure guarantees compactness in the public key.
In such a context, the additional geometrical structure can be added without exposing the secret code: \ac{QC-LDPC} and \ac{QC-MDPC} code-based cryptosystems have been extensively studied in recent years~\cite{Baldi2008, Misoczki2013, Baldi2018, LEDAcrypt, BIKE2017} and currently achieve relatively very small public keys.
However, differently from the bounded-distance decoders used for algebraic codes (like the mentioned Goppa codes), the iterative decoders used for sparse parity-check codes are not characterized by a deterministic decoding radius and, thus, decoding might fail with some probability, or \ac{DFR}.

Such a feature is crucial, since it has been shown how this probabilistic nature of the decoder actually exposes the system to cryptanalysis techniques based on the observation of the decryption phase.
State-of-the-art attacks of this kind are commonly called reaction attacks, when based on decoding failures events~\cite{Guo2016, Fabsic2017, Fabsic2018, Paiva2018}, or 
side-channel attacks, when based on information such as the duration of the decoding phase (in this case we speak properly of timing attacks) or other quantities~\cite{Fabsic2016, Eaton2018, Nilsson2018}.
All these previous techniques exploit the \ac{QC} structure of the code and aim at recovering some characteristics of the secret key by performing a statistical analysis on a sufficiently large amount of collected data.
The rationale is that many quantities that are typical of the decryption procedure depend on
a statistical relation between some properties of the secret key and the error vector that is used during encryption.
Thus, after observing a sufficiently large number of decryption instances, an adversary can exploit the gathered information to reconstruct the secret key, or an equivalent version of it.
The reconstruction phase is commonly very efficient, unless some specific choices in the system design are made~\cite{Santini2018ISIT, Santini2018CANS} which, however, may have some significant drawbacks in terms of public key size.
All the aforementioned attack techniques are instead prevented if the \ac{DFR} has negligible values \cite{Tillich2018} and the algorithm is implemented with constant time.
Nevertheless, at their current state, these solutions are far from being practical and efficient, and the use of ephemeral keys (which means that each key-pair is refreshed after just one decryption) is necessary to make these systems secure~\cite{BIKE2017, Baldi2018}.


In this paper we study reaction and timing attacks, and we show that information leakage in the decoding phase can actually be related to the number of overlapping ones between columns of the secret parity-check matrix.
Furthermore, we show that all attacks of this kind can be analyzed as a unique procedure, which can be applied to recover information about the secret key, regardless of the code structure.
Such an algorithm, when applied on a \ac{QC} code, permits to recover an amount of information greater than that retrievable through previously published attacks.
Moreover, we provide an approximate model that allows predicting the behaviour of the decoder in the first iteration with good accuracy.
This model justifies the phenomenon that is at the basis of all the aforementioned attacks and can be even used to conjecture new attacks. 
Our results are confirmed by numerical simulations and enforce the employment of constant time decoders, with constant power consumption and negligible \ac{DFR}, in order to allow for the use of long-lasting keys in these systems.

The paper is organized as follows: Section \ref{sec:Preliminaries} describes the notation used throughout the manuscript and provides some basic notions about cryptosystems based on sparse parity-check codes.
In Section \ref{sec:Framework} we summarize state-of-the-art reaction and timing attacks, and present a general algorithm that can be used to attack any sparse parity-check code. 
An approximate model for the analysis of the first iteration of the BF decoder is presented in 
Section \ref{sec:model}.
In Section \ref{sec:NewSideChannel} we describe some additional sources of information leakage, that can be used by an opponent to mount a complete cryptanalysis of the system.
Finally, in Section \ref{sec:Conclusion} we draw some conclusive remarks.

\section{Preliminaries\label{sec:Preliminaries}}
We represent matrices and vectors through bold capital and small letters, respectively.
Given a matrix $\bA$, we denote as $\ba_i$ its $i$-th column and as  $a_{i,j}$ its element in position $(i,j)$. 
Given a vector $\bb$, its $i$-th entry is referred to as $b_i$; its support is denoted as $\phi(\bb)$ and is defined as the set containing the positions of its non-null entries.
The vector $\bzero_n$ corresponds to the all-zero $n$-tuple; the function returning the Hamming weight of its input is denoted as $\mathrm{wt}\{\cdot\}$.

\subsection{LDPC and MDPC code-based cryptosystems\label{sec:LDPC}}

The schemes we consider are built upon a code $\mathcal{C}$ described by a sparse parity-check matrix $\bH\in\mathbb{F}_2^{r\times n}$, where $n$ is the code blocklength.
We here focus on the case of regular matrices, in which all the rows and all the columns have constant weights respectively equal to $w\ll n$ and $v\ll r$.
The code $\mathcal{C}$ is then  $(v,w)$-regular, and is commonly called \ac{LDPC} code if $w=O(\log{n})$, or \ac{MDPC} code, if $w=O(\sqrt{n})$. 
Regardless of such a distinction, these two families of codes actually have similar properties: they can be decoded with the same decoding algorithms and are thus exposed in the same way to the attacks we consider.
So, from now on we will not distinguish between these two families, and just refer to $(v,w)$-regular codes.

In the McEliece framework, the public key is a generator matrix $\bG$ for $\mathcal{C}$; a ciphertext is obtained as
\begin{equation}
    \bx = \bm \bG + \be,
\end{equation}
where $\bm\in\mathbb{F}_2^{k}$ is the plaintext and $\be$ is a randomly generated $n$-tuple with weight $t$.
Decryption starts with the computation of the syndrome $\bs = \bH\bx^T = \bH \be^T$, where $^T$ denotes transposition.
Then, an efficient syndrome decoding algorithm is applied on $\bs$, in order to recover $\be$.

In the Niederreiter formulation, the public key is a parity-check matrix $\bH'=\bS\bH$ for $\mathcal{C}$, where $\bS$ is a dense non singular matrix.
The plaintext $\bm$ is converted into an $n$-tuple $\be$ with weight $t$ by means of an invertible mapping

\begin{equation}
    \mathcal{M} \mapsto \{\be \in \mathbb{F}_2^n| \mathrm{wt}\{\be\}=t \},
    \label{eq:mapping}
\end{equation}
where $\mathcal{M}$ is the space of all possible plaintexts $\bm$.
The ciphertext is then computed as
\begin{equation}
    \bx=\bH'\be^T.
\end{equation}

Decryption starts with the computation of $\bs = \bS^{-1}\bx$; then, an efficient syndrome decoding algorithm is applied on $\bs$, in order to recover $\be$, from which the plaintext $\bm$ is reconstructed by inverting \eqref{eq:mapping}.

Regardless of the particular system formulation we are considering (McEliece or Niederreiter), the decryption phase relies on a syndrome decoding algorithm, applied on the syndrome of a weight-$t$ error vector. 
Since, in the attacks we consider, information is leaked during the decoding phase, we will not distinguish between the McEliece and the Niederreiter formulation in the following.

The decoding algorithm must show a good trade-off between complexity and \ac{DFR}; for this reason, a common approach is that of relying on the so-called \ac{BF} decoders, whose principle has been introduced by Gallager \cite{Gallager1963}. 
The description of a basic \ac{BF} decoding procedure is given in Algorithm \ref{alg:BF}.
\begin{algorithm}[ht!]
\caption{{\fontfamily{cmss}\selectfont BFdecoder}}\label{alg:BF}
\hspace*{\algorithmicindent} \textbf{Input:}  $\bH\in\mathbb{F}_2^{r\times n}$, $\bs\in\mathbb{F}_2^r$, $i_{\texttt{max}}\in\mathbb{N}$, $b \in\mathbb{N}$\\
\hspace*{\algorithmicindent}\textbf{Output}: $\be' \in \mathbb{F}_2^n$, $\bot\in\mathbb{F}_2$
\begin{algorithmic}[1]
\State{$\be'\gets \bzero_{n}$}
\State{$\bot\gets 0$}
\State{$i\gets 0$}
\While{$\mathrm{wt}\{\bs\}>0 \wedge i< i_{\texttt{max}}$}
\State{$\Psi\gets \varnothing$}
\For{$j\gets 0\hspace{2mm}\textbf{to}\hspace{2mm}n-1$}
\State{$\sigma_j\gets 0$}
\For{$l\in\phi(\bh_j)$}
\State{$\sigma_j\gets \sigma_j+s_l$}
\EndFor
\If{$\sigma_j \geq b$}
\State{$\Psi\gets \Psi \cup j$}\Comment{Position $j$ is estimated as error affected}
\EndIf
\EndFor
\For{$j\in\Psi$}
\State{$e'_j\gets \neg e'_j$}\Comment{Error estimation update}
\State{$\bs\gets \bs+\bh_j$}\Comment{Syndrome update}
\EndFor
\State{$i\gets i+1$}
\EndWhile
\If{$\mathrm{wt}\{\bs\}>0$}
\State{$\bot \gets 1$}\Comment{Decoding failure}
\EndIf
\\\Return{$\{\be',\bot\}$}
\end{algorithmic}
\end{algorithm}
The decoder goes through a maximum number of iterations $i_{\texttt{max}}$, and at each iteration it exploits a likelihood criterion to estimate the error vector $\be$. Outputs of the decoder are the estimate of the error vector $\be'$ and a boolean value $\bot$ reporting events of decoding failure.
When $\bot=0$, we have $\be'=\be$, and decoding was successful; if $\bot=1$, then $\be'\neq \be$ and we have encountered a decoding failure. So, clearly, the \ac{DFR} can be expressed as the probability that $\bot=1$, noted as $P\{\bot=1\}$. The likelihood criterion is based on a threshold $b$ (line 11 of the algorithm), which, in principle, might also vary during the iterations (for instance, some possibilities are discussed in \cite{Misoczki2013}); all the simulations results we show in this paper are referred to the simple case in which the threshold is kept constant throughout all the decoding procedure.
In particular, in the simulations we have run, the values of $i_{\texttt{max}}$ and $b$ have been chosen empirically. 
Our analysis is general and can be easily extended to other decoders than those considered here.
Indeed, many different decoders have been analyzed in the literature (for instance, see \cite{Eaton2018} and \cite{Nilsson2018}), and, as for the outcome of reaction and timing attacks, there is no meaningful difference between them.
This strongly hints that such attacks are possible because of the probabilistic nature of the decoder, and are only slightly affected by the particular choice of the decoder and its settings.
However, the analysis we provide in Section \ref{sec:model}, which describes the decoder behaviour in the first iteration, takes into account the effect of the threshold value.

We point out that Algorithm \ref{alg:BF} is commonly called an \emph{out-of-place} decoder, as the syndrome $\bs$ is updated after the set $\Psi$ is computed.
A different procedure is the one of \emph{in-place} decoders, in which the syndrome is updated every time a bit is estimated as error affected (i.e., after the $\it if$ instruction in line 11). In this paper we only focus on \emph{out-of-place} decoders. The reason is that the attacks we consider seem to be emphasized when \emph{in-place} decoders are used\cite{Eaton2018, Nilsson2018}. However, even if a careful analysis is needed, it is very likely that our results can be extended also to \emph{in-place} decoders.

\section{A general framework for reaction and timing attacks\label{sec:Framework}}
In this section we describe a family of attacks based on statistical analyses, namely \emph{statistical attacks}. This family includes reaction attacks, in which data is collected through the observation of Bob's reactions, and side-channel attacks. A statistical attack of the types here considered can be described as follows.

Let us consider a public-key cryptosystem with private and public keys $K_S $ and $K_P$, respectively, and security parameter $\lambda$ (i.e., the best attack on the system has complexity $>2^{\lambda}$). 
We denote as {\fontfamily{cmss}\selectfont Decrypt$(K_S,\bx)$} a decryption algorithm that, given a ciphertext $\bx$ and $K_S$ as inputs, returns either the plaintext $\bm$ or a decryption failure.
We define $\mathcal{D}(K_S,\bx)$ as an oracle that, queried with a ciphertext $\bx$, runs {\fontfamily{cmss}\selectfont Decrypt$(K_S,\bx)$} and returns some metrics that describe the execution of the decryption algorithm.
More details about the oracle's replies are provided next.
An adversary, which is given $K_P$, queries the oracle with $N$ ciphertexts $\{\bx^{(i)}\left| i=1,\cdots,N\right.\}$; we denote as $y^{(i)}$ the oracle's reply to the $i$-th query $\bx^{(i)}$.
The adversary then runs an algorithm $\mathcal{A}(K_P,\{\bx^{(0)},y^{(0)}\},\cdots,\{\bx^{(N-1)},y^{(N-1)}\})$ that takes as inputs $K_P$ and the pairs of oracle queries and replies, and returns $K'_S$. 
The algorithm $\mathcal{A}$ models the procedure that performs a statistical analysis of the gathered data and reconstructs the secret key, or an equivalent version of it.
The time complexity of this whole procedure can be approximated as 
\begin{equation}
C = \alpha N + C_{\mathcal{A}},
\end{equation} 
where $\alpha$ corresponds to the average number of operations performed for each query and $C_{\mathcal{A}}$ is the complexity of executing the algorithm $\mathcal{A}$. 
The adversary is then challenged with a randomly generated ciphertext $\bx^*$, corresponding to a plaintext $\bm^*$.
We consider the attack successful 
if $C<2^{\lambda}$ and the probability of $\bm =$ {\fontfamily{cmss}\selectfont Decrypt$(K'_S,\bx^*)$} being equal to $\bm^*$ is not negligible (i.e., larger than $2^{-\lambda}$).

We point out that this formulation is general, since it does not distinguish between the McEliece and Niederreiter cases. In the same way the private and public keys might be generic. For example, this model describes also reaction attacks against LEDA cryptosystems ~\cite{Baldi2018}, in which the secret key consists of $\bH$ and an additional sparse matrix $\mathbf{Q}$.

The above model allows for taking into account many kinds of attacks, depending on the oracle's reply.
For instance, when considering attacks based on decryption failures, the oracle's reply is a boolean value which is true in case of a failure and false otherwise.
When considering timing attacks based on the number of iterations, then the oracle's reply corresponds to the number of iterations run by the decoding algorithm.

In this paper we focus on systems with security against a \ac{CCA}, that is, the case in which a proper conversion (like the one of ~\cite{Kobara2001}) is applied to the McEliece/Niederreiter cryptosystem, in order to achieve \ac{CCA} security. 
In our attack model, this corresponds to assuming that the oracle queries are all randomly generated, i.e., the error vectors used during encryption can be seen as randomly picked elements from the ensemble of all $n$-uples with weight $t$.
Opposed to the \ac{CCA} case, in the \ac{CPA} case the opponent is free to choose the error vectors used during encryption: from the adversary standpoint, the \ac{CPA} assumption is clearly more optimistic than that of \ac{CCA}, and leads to improvements in the attack~\cite{Guo2016, Nilsson2018}. 
Obviously, all results we discuss in this paper can be extended to the \ac{CPA} case.

One final remark is about the schemes we consider: as shown in \cite{Santini2018ISIT,Santini2018CANS}, the complexity of algorithm $\mathcal{A}$ can be increased with proper choices in the structure of the secret key.
Basically, in these cases the adversary can gather information about the secret key, but cannot efficiently use this information to reconstruct the secret key, or to obtain an equivalent version of it.
In this paper we do not consider such approaches and we assume that the algorithm $\mathcal{A}$ always runs in a feasible time, as it occurs in~\cite{Guo2016, Fabsic2017}. 

\subsection{State-of-the-art statistical attacks}
Modern statistical attacks \cite{Guo2016, Eaton2018, Fabsic2017, Fabsic2018, Nilsson2018} are specific to the sole case of \ac{QC} codes having the structure originally proposed in \cite{Baldi2007ISIT}, which are defined through a secret parity-check matrix in the form 
\begin{equation}
\label{eq:privateH}
    \bH=\begin{bmatrix}\bH_0,\bH_1,\ldots,\bH_{n_0-1}\end{bmatrix},
\end{equation}
where each $\bH_i$ is a circulant matrix of weight $v$ and $n_0$ is a small integer.
Thus, the corresponding code is a $(v,n_0v)$-regular code.

All existing statistical attacks are focused on guessing the existence (or absence) of some cyclic distances between symbols $1$ in $\bH$.
In particular, an adversary aims at recovering the following quantities, which were introduced in~\cite{Guo2016}.
\vspace{2mm}
\\\textbf{Distance spectrum: }\emph{
 Given a vector $\ba$, with support $\phi(\ba)$ and length $p$,  its \emph{distance spectrum} is defined as
\begin{equation}
\label{eq:ds}
    \mathrm{DS}(\ba) =\left\{\left.\min\{\pm(i-j)\mod{p}\}\right|i,j\in\phi(\ba),\hspace{2mm}i\neq j\right\}.
\end{equation}
}
\\\hspace{2mm}
\textbf{Multiplicity: }\emph{
We say that a distance $d\in \mathrm{DS}(\ba)$ has \emph{multiplicity} $\mu_d$ if there are $\mu_d$ distinct pairs in $\phi(\ba)$ which produce the same distance $d$.
}
\\\hspace{2mm}
\\Basically, the distance spectrum is the set of all distances with multiplicity larger than $0$.
It can be easily shown that all the rows of a circulant matrix are characterized by the same distance spectrum; thus, given a circulant matrix $\mathbf M$, we denote the distance spectrum of any of its rows (say, the first one) as $\mathrm{DS}(\mathbf M)$.

Statistical attacks proposed in the literature aim at estimating the distance spectrum of the circulant blocks in the secret $\bH$, and are based on the observation that some quantities that are typical of the decryption procedure depend on the number of common distances between the error vector and the rows of the parity-check matrix.  
In particular, the generic procedure of a statistical attack on a cryptosystem whose secret key is in the form \eqref{eq:privateH} is described in Algorithm \ref{alg:ExGJS}; we have called the algorithm {\fontfamily{cmss}\selectfont Ex-GJS} in order to emphasize the fact that it is an extended version of the original GJS attack~\cite{Guo2016}, which was only focused on a single circulant block in $\bH$. Our algorithm, which is inspired by that of~\cite{Fabsic2017}, is a generalization of the procedure in~\cite{Guo2016}, in which all the circulant blocks in $\bH$ are taken into account. 
We present this algorithm in order to show the maximum amount of information that state-of-the-art statistical attacks allow to recover.
\begin{algorithm}[ht!]
\caption{{\fontfamily{cmss}\selectfont Ex-GJS}}\label{alg:ExGJS}
\hspace*{\algorithmicindent} \textbf{Input:} public key $K_P$, number of queries $N\in\mathbb{N}$\\
\hspace*{\algorithmicindent} \textbf{Output}: estimates $\ba^{(0)}, \cdots, \ba^{(n_0-1)}$, $\bb^{(0)}, \cdots, \bb^{(n_0-1)}\in\mathbb{N}_{\left\lfloor p/2 \right\rfloor}$.
\begin{algorithmic}[1]
\State{Initialize $\ba^{(0)}, \cdots, \ba^{(n_0-1)}$, $\bb^{(0)}, \cdots, \bb^{(n_0-1)}$ as null arrays of length $\left\lfloor p/2\right\rfloor$}
\For{$i\gets 1\hspace{2mm}\textbf{to}\hspace{2mm}N$}
    \State{$\bx^{(i)}\gets$ ciphertexts obtained through the error vector $\be^{(i)}$}
    \State{$y^{(i)} \gets \mathcal{D}(K_S,\bx^{(i)})$}
    \For{$j\gets 0\hspace{2mm}\textbf{to}\hspace{2mm}n_0-1$}
        \State{$\Delta_j\gets \mathrm{DS}(\be^{(i)}_j)$}
        \For{$d\in\Delta_j$}
            \State{$a^{(j)}_d\gets a^{(j)}_d+y^{(i)}$}
            \State{$b^{(j)}_d\gets b^{(j)}_d+1$}
        \EndFor
    \EndFor
\EndFor
\\\Return{$\{\ba^{(0)}, \cdots, \ba^{(n_0-1)}$, $\bb^{(0)}, \cdots, \bb^{(n_0-1)}\}$}
\end{algorithmic}
\end{algorithm}

The error vector used for the $i$-th query is expressed as
$\be^{(i)}=[\be^{(i)}_0,\ldots, \be^{(i)}_{n_0-1}]$, where each $\be^{(i)}_j$ has length $p$. The estimates $\ba^{(0)}, \ldots, \ba^{(n_0-1)}$ and $\bb^{(0)}, \ldots, \bb^{(n_0-1)}$ are then used by the adversary to guess the distance spectra of the blocks in the secret key.
Indeed, let us define  $\mathcal{E}^{(d,j)}(n,t)$ as the ensemble of all error vectors having length $n$, weight $t$ and such that they exhibit a distance $d$ in the distance spectrum of the $j$-th length-$p$ block.
Then, depending on the meaning of the oracle's reply, the ratios $a^{(j)}_{d} / b^{(j)}_d$ correspond to the estimate of the average value of some quantity, when the error vector belongs to $\mathcal{E}^{(d,j)}(n,t)$.
For instance, when considering attacks based on decryption failures, the oracle's reply is either $0$ or $1$, depending on whether the decryption was successful or failed.
In such a case, the ratio $a^{(j)}_{d} / b^{(j)}_d$ corresponds to the empirical measurement of the \ac{DFR}, conditioned to the event that the error vector belongs to $\mathcal{E}^{(j,d)}(n,t)$.
In general, statistical attacks are successful because many quantities that are typical of the decoding procedure depend on the multiplicity of the distances in $\mathrm{DS}(\bH_j)$.
In the next section we generalize this procedure, by considering different ensembles for the error vectors; 
 then, in Section \ref{sec:model}, we provide a theoretical explanation for such a phenomenon.

\subsection{Exploiting decryption failures on generic codes}
In this section we generalize the {{\fontfamily{cmss}\selectfont Ex-GJS}} procedure, and describe an algorithm which can be used to recover information about any regular code.
In particular, our analysis shows that events of decoding failure i) do not strictly depend on the \ac{QC} structure of the adopted code, and
ii) permit to retrieve a quantity that is more general than distance spectra. 

\begin{figure}
    \centering
    \includegraphics[keepaspectratio, width = 9cm]{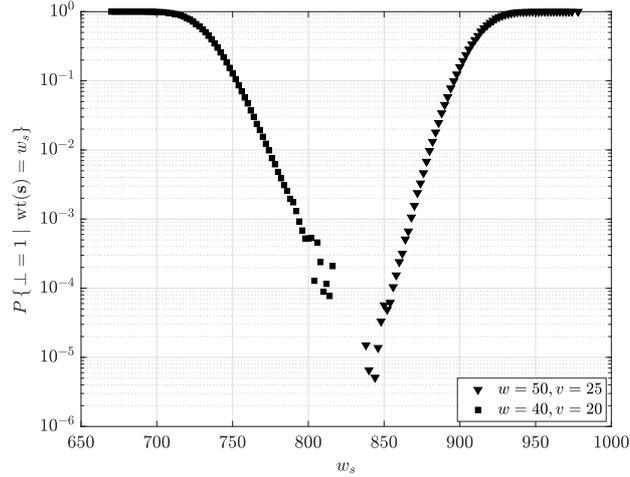}
    \caption{Distribution of the \ac{DFR} as a function of the syndrome weight, for two regular $(v,w)$-regular \ac{LDPC} codes, decoded through Algorithm \ref{alg:BF} with $i_{\texttt{max}}=5$ and $b=15$. The weight of the error vectors is $t=58$; for each code, $10^7$ decoding instances have been considered.} 
    \label{fig:ws_vs_dfr}
\end{figure}

We first show that, for generic regular codes, there  is a connection between the syndrome weight and the \ac{DFR}. This statement is validated by numerical simulations on $(v,w)$-regular codes, obtained through Gallager construction~\cite{Gallager1963}, in which $\frac{v}{w}=\frac{r}{n}$. In particular, we have considered two codes with length $n=5000$, redundancy $r=2500$ and different pairs $(v,w)$, decoded through Algorithm \ref{alg:BF}; their \ac{DFR} (i.e., the probability of Algorithm \ref{alg:BF} returning $\bot=1$) vs.  syndrome weight is shown in Fig.~\ref{fig:ws_vs_dfr}.   
We notice from Fig.~\ref{fig:ws_vs_dfr} that there is a strong dependence between the initial syndrome weight and the \ac{DFR} and that different pairs $(v,w)$ can lead to two different trends in the \ac{DFR} evolution. Section \ref{sec:model} is devoted to the explanation of this phenomenon.

Let us now define $\mathcal{E}(n,t,i_0,i_1)$ as the ensemble of all vectors having length $n$, weight $t$ and whose support contains elements $i_0$ and $i_1$.
Let $\bs$ be the syndrome of an error vector $\be\in\mathcal{E}(n,t,i_0,i_1)$: we have 
\begin{equation}
    \bs = \bh_{i_0}+\bh_{i_1}+\sum_{j\in\phi(\be)\setminus\{i_0,i_1\}}{\bh_j}.
\end{equation}
The syndrome weight has a probability distribution that depends on the interplay between $\bh_{i_0}$ and $\bh_{i_1}$: basically, when these two columns overlap in a small (large) number of ones, then the average syndrome weight gets larger (lower). 
Moreover, motivated by the empirical evidence of Fig. \ref{fig:ws_vs_dfr}, one can expect that the \ac{DFR} experienced over error vectors belonging to different ensembles $\mathcal{E}(n,t,i_0,i_1)$ depends on the number of overlapping ones between columns $\bh_{i_0}$ and $\bh_{i_1}$.
Then, a statistical attack against a generic regular code can be mounted, as described in Algorithm \ref{alg:generic_reaction}, which we denote as General Statistical Attack (GSA). The output of the algorithm is represented by the matrices $\bA$ and $\bB$, which are used by the adversary to estimate the average value of the oracle's replies, as a function of the pair $(i_0,i_1)$.  
Notice that in Algorithm \ref{alg:generic_reaction} the oracle's reply is denoted as $y^{(i)}$ and does not need to be better specified. We will indeed show in Section \ref{sec:NewSideChannel} that the same procedure can be used to exploit other information sources than the success (or failure) of decryption.

We now focus on the case of $y^{(i)}$ being $0$ or $1$, depending on whether decryption was successful or not.
Then, each ratio $a_{j,l}/b_{j,l}$ is the empirical estimate of the probability of encountering a decryption failure, when the error vector contains both $j$ and $l$ in its support.
\begin{algorithm}[ht!]
\caption{{\fontfamily{cmss}\selectfont GSA }}\label{alg:generic_reaction}
\hspace*{\algorithmicindent} \textbf{Input:} public key $K_P$, number of queries $N\in\mathbb{N}$\\ 
\hspace*{\algorithmicindent} \textbf{Output}: estimates $\bA,\bB\in\mathbb{N}^{n\times n}$.
\begin{algorithmic}[1]
   \State{$\bA\gets \mathbf 0_{n\times n}$}
    \State{$\bB\gets \mathbf 0_{n\times n}$}
\For{$i\gets 1\hspace{2mm}\textbf{to}\hspace{2mm}N$}
    \State{$\bx^{(i)}\gets$ ciphertexts obtained through the error vector $\be^{(i)}$}
    \State{$y^{(i)} \gets \mathcal{D}(K_S,\bx^{(i)})$}
    \State{$\phi(\be^{(i)})\gets$ support of $\be^{(i)}$}
    \For{$j\in\phi(\be^{(i)})$}
        \For{$l\in\phi(\be^{(i)})$}
            \State{$a_{j,l}\gets a_{j,l} + y^{(i)}$}
            \State{$b_{j,l}\gets b_{j,l} + 1$}
        \EndFor
    \EndFor
\EndFor
\\\Return{$\{\bA,\bB\}$}
\end{algorithmic}
\end{algorithm}
\begin{figure}%
\centering
\subfigure[]{\includegraphics[keepaspectratio, width = 5.5cm]{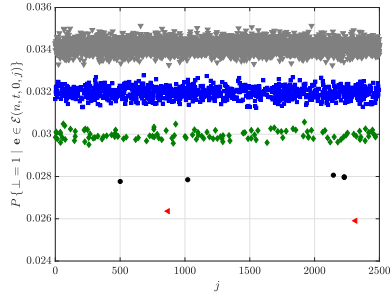}}\qquad
\subfigure[]{\includegraphics[keepaspectratio, width = 5.5cm]{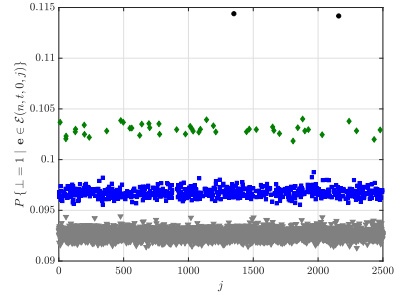}}\\
\caption{Simulation results for  $(v,w)$-regular code with $n=5000$, $k=2500$, for $t=58$ and for error vector belonging to ensembles $\mathcal{E}(n,t,0,j)$, for $j\in [1,\ldots,n-1]$. The parameters of the codes are $v=25$, $w=50$ for Figure (a), $v=20$, $w=40$ for Figure (b); the decoder settings are $i_{\texttt{max}}=5$ and $b=15$. The results have been obtained through the simulation of $10^9$ decoding instances.
Grey, blue, green, black and red markers are referred to pairs of columns with number of intersections equal to $0,1,2,3,4$, respectively.}
\label{fig:subfigurew50_W40}
\end{figure}
One might expect that the ratios $a_{j,l} / b_{j,l}$ are distributed on the basis of the number of overlapping ones between columns $j$ and $l$ in $\bH$. 
We have verified this intuition by means of numerical simulations; the results we have obtained are shown in Fig. \ref{fig:subfigurew50_W40}, for the case of error vectors belonging to ensembles $\mathcal{E}(n,t,0,j)$, with $j\in[1,\ldots, n-1]$.
The figure clearly shows that the ratios $a_{j,l}/b_{j,l}$ can be used to guess the number of overlapping ones between any pair of columns in $\bH$. 

These empirical results confirm the conjecture that the \ac{DFR} corresponding to error vectors in $\mathcal E(n,t,i_0,i_1)$ depends on the number of overlapping ones between the columns $i_0$ and $i_1$. Moreover, these results show that the same idea of~\cite{Guo2016}, with some generalization, can be applied to whichever kind of code.

We now show that even when \ac{QC} codes are considered, our algorithm recovers more information than that which can be obtained through the {\fontfamily{cmss}\selectfont Ex-GJS} procedure. For such a purpose, let us consider a parity-check matrix in the form \eqref{eq:privateH}, and let $\gamma_{i,j}$ be the number of overlapping ones between columns $i$ and $j$. 
Now, because of the \ac{QC} structure, we have
\begin{equation}
\label{eq:qc_lambda_ij}
    \left|\bh_i \cap\bh_j \right| = \left|\bh_{p\left\lfloor i/p \right\rfloor+[i+z\modp]} \cap\bh_{p\left\lfloor j/p \right\rfloor+[j+z\modp]} \right|,\hspace{2mm}\forall z.
\end{equation}

We now consider two columns that belong to the same circulant block in $\bH$, i.e. $i=pi_p+i'$, $j=pi_p+j'$, where $0\leq i_p \leq n_0-1$; then, \eqref{eq:qc_lambda_ij} can be rewritten as
\begin{equation}
    \left|\bh_i \cap\bh_j \right| = \left|\bh_{pi_p+[i'+z\modp]} \cap\bh_{pi_p+[j'+z\modp]} \right|,\hspace{2mm}\forall z.
\end{equation}
With some simple computations, we finally obtain
\begin{equation}
\label{eq:lambda_rule}
    \left|\bh_{pi_p+i'}\cap \bh_{pi_p+j'}\right| =  \begin{cases}\left|\bh_{pi_p} \cap\bh_{pi_p+p-(i'-j')}\right| &\text{if $j'<i'$}\\
    \left|\bh_{pi_p} \cap\bh_{pi_p+j'-i')}\right| &\text{if $j'>i'$}
    \end{cases},
\end{equation}
which holds for all indices $ i',j'\in[0,\ldots,p-1],i'\neq j'$. Similar considerations can be carried out if the two columns do not belong to the same circulant block. So, \eqref{eq:lambda_rule} shows that the whole information about overlapping ones between columns in $\bH$ is actually represented by a subset of all the possible values of $\gamma_{i,j}$.
This means that the execution of Algorithm \ref{alg:generic_reaction} can be sped up by taking the \ac{QC} structure into account.

In particular, the values of $\gamma_{i,j}$ can be used to obtain the distance spectra of the blocks in $\bH$ in a straightforward way.
Let us refer to Equation \eqref{eq:lambda_rule}, and look at two columns $\bh_{pi_p}$ and $\bh_{j}$, with $j=pi_p+j'$, where $j'\in[0,1,\ldots,p-1]$. 
We denote the support of $\bh_{pi_p}$ as $\phi(\bh_{pi_p})=\{c^{(pi_p)}_0,\cdots,c^{(pi_p)}_{v-1}\}$.
The support of $\bh_j$ can be expressed as
\begin{equation}
    \phi(\bh_j)=\{c^{(j)}_l\left| c^{(j)}_l = c^{(pi_p)}_l+j'\modp,\hspace{2mm} l\in[0,\ldots,v-1],\hspace{2mm}c^{(pi_p)}_l\in\phi(\bh_{pi_p})\right.\}.
\end{equation}
Then, we have $\left| \phi(\bh_{pi_p})\cap\phi(\bh_j)\right|=\gamma_{pi_p,j}$; this means that there are $\gamma_{pi_p,j}$ pairs $\{c,c'\}\in\phi(\bh_{pi_p})\times \phi(\bh_j)$ such that
\begin{equation}
\label{eq:apr_a}
    c' = c + d \modp, \hspace{2mm} d \in\{j',p-j'\}.
\end{equation}
It is easy to see that \eqref{eq:apr_a} corresponds to the definition of the distance spectrum of the blocks in $\bH$; then, \eqref{eq:apr_a} can be turned into the following rule 
\begin{equation}
\left| \phi(\bh_{pi_p})\cap \phi(\bh_pi_p+j')\right|=\gamma \leftrightarrow d \in  \mathrm{DS}(\bH_{i_p}), \hspace{2mm}\mu_{d}=\gamma,\end{equation}
with $d=\min\{\pm j' \modp\}$.

This proves that the procedure described by Algorithm \ref{alg:generic_reaction} allows obtaining at least the same amount of information recovered through the {\fontfamily{cmss}\selectfont Ex-GJS } algorithm, which is specific to the \ac{QC} case and guarantees a complete cryptanalysis of the system~\cite{Guo2016}.
In other words, our analysis confirms that Algorithm \ref{alg:generic_reaction} is applicable and successful in at least all the scenarios in which the attack from~\cite{Guo2016} works. Moreover, our procedure allows for recovering a larger amount of information about the secret key, and thus defines a broader perimeter of information retrieval, which encompasses existing and future attacks. 

\section{An approximated model for reaction attacks\label{sec:model}}
\begin{figure}%
\centering
\subfigure[]{\includegraphics[keepaspectratio, width = 5.5cm]{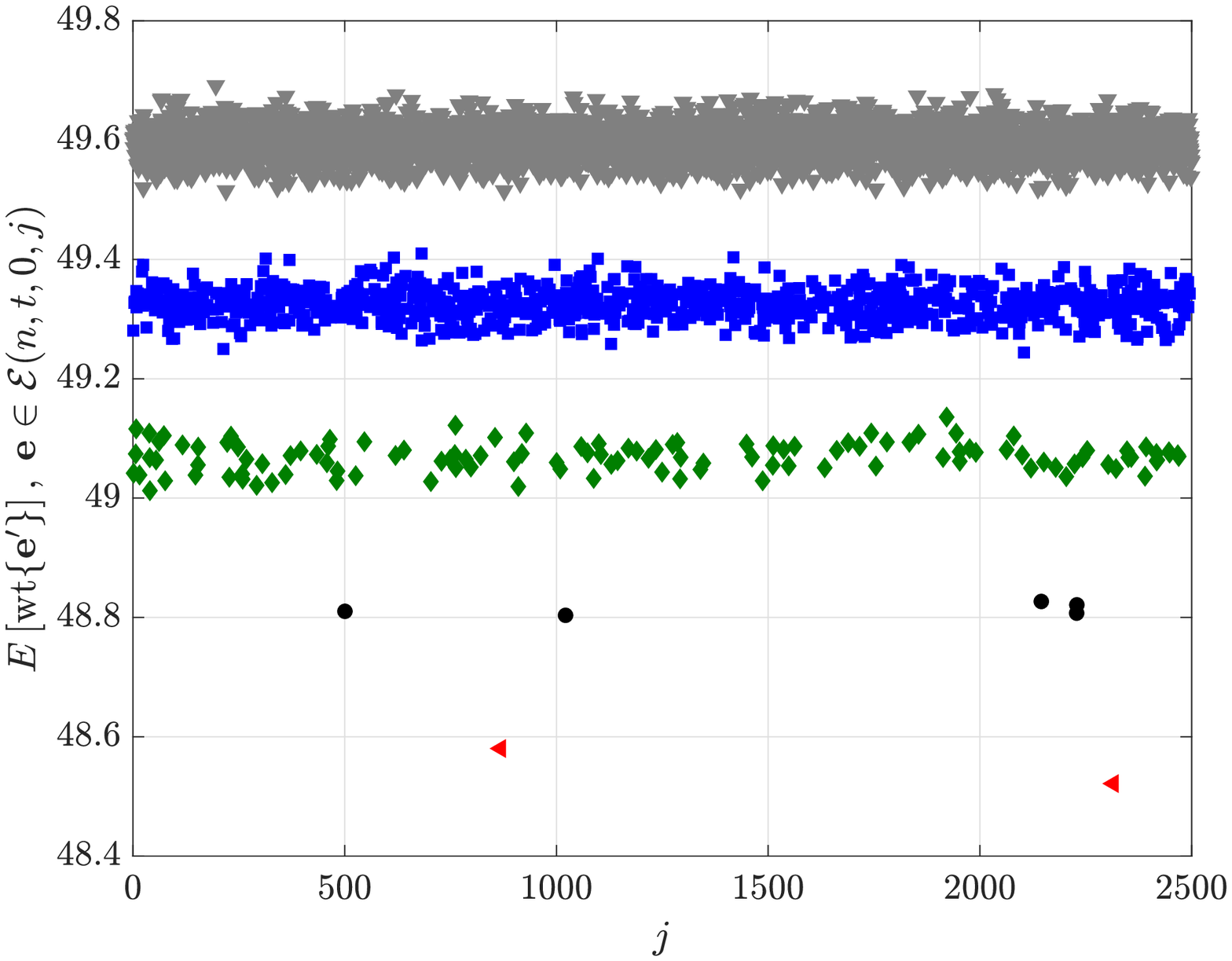}}\qquad
\subfigure[]{\includegraphics[keepaspectratio, width = 5.5cm]{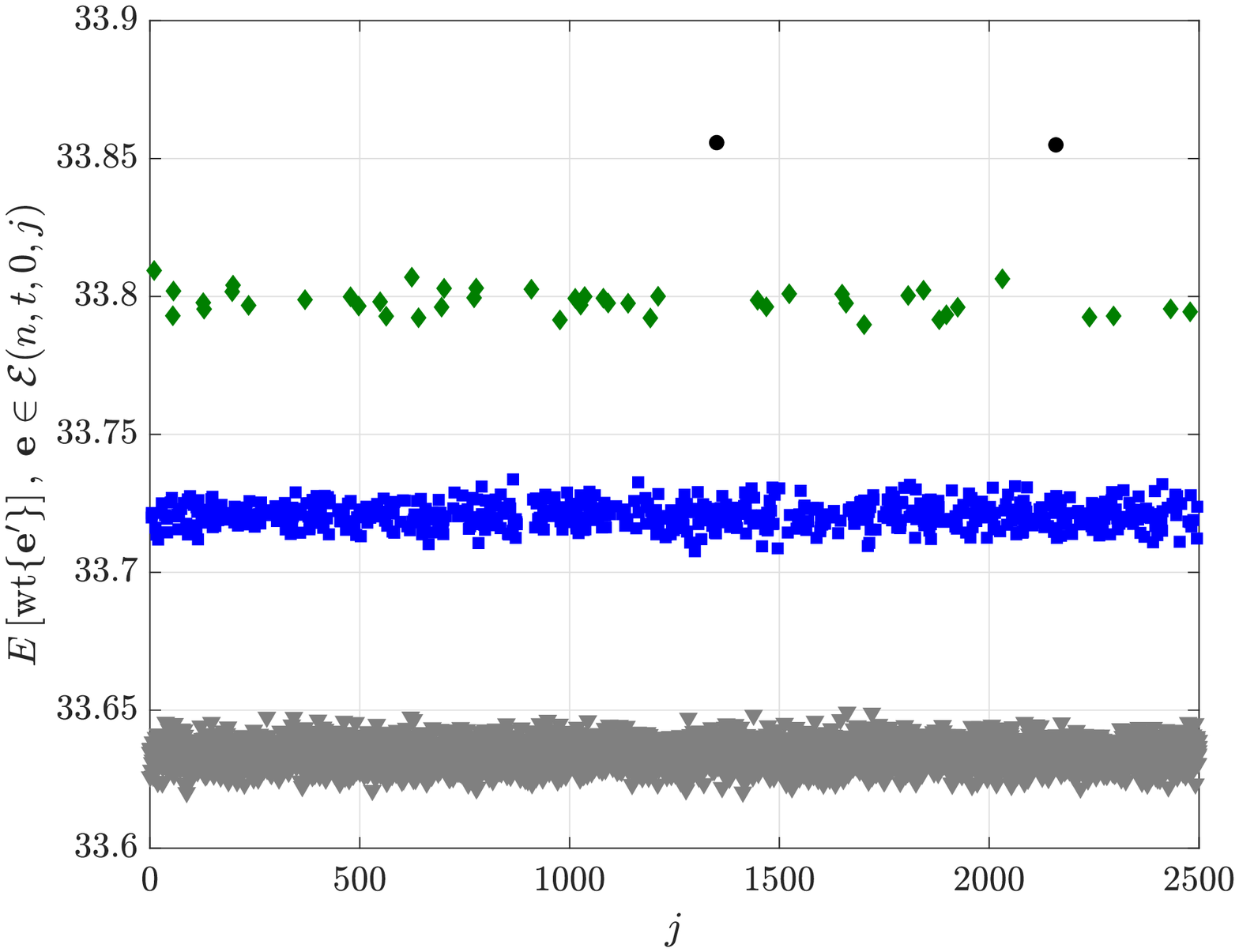}}\\
\caption{Simulation results for the number of residual errors after the first iteration of \ac{BF} decoding, for  $(v,w)$-regular codes with $n=5000$, $k=2500$, for $t=58$ and for error vector drawn from ensembles $\mathcal{E}(n,t,0,j)$, for $j\in [1,\ldots,n-1]$. The parameters of the codes are $v=25$, $w=50$ for Figure (a), $v=20$, $w=40$ for Figure (b); the decoder settings are $i_{\mathrm{max}}=5$ and $b=15$. The results have been obtained through the simulation of $10^9$ decoding instances. 
Grey, blue, green, black and red markers are referred to pairs of columns with number of intersections equal to $0,1,2,3,4$, respectively. }
\label{fig:num_errors_w50_w40}
\end{figure}

The main result of this section is summarized in the following proposition, for which we 
provide theoretical justifications and empirical evidences.
\begin{Pro}
Let $\bH$ be the parity-check matrix of a $(v,w)$-regular code, which is decoded through 
Algorithm \ref{alg:BF} with decoding threshold $b$.
Let $(i_0,i_1)$ and $(i^*_0,i^*_1)$ be two distinct pairs of indexes, and consider error vectors 
$\be\in\mathcal{E}(n,t,i_0,i_1)$ and $\be^*\in\mathcal{E}(n,t,i^*_0,i^*_1)$.
Let $\epsilon$ and $\epsilon^*$ be the probabilities that $\be$ and $\be^*$ result in a decoding failure, respectively.
Let $\be'_{[1]}$ be the error vector estimate after the first iteration; we define $\be' = \be + \be'_{[1]}$ and $t'= \mathrm{wt}\{\be'\}$.
Then,  $\epsilon > \epsilon^*$ if and only if $E\left[t'\right] >  E\left[t'^{*}\right]$, where $E\left[\hspace{1mm}\cdot\hspace{1mm}\right]$ denotes the expected value.
\label{pro:DFR_t_prime}
\end{Pro}
Essentially, the above proposition implies that increments (reductions) of the \ac{DFR} are due to the fact that, 
depending on the particular matrix $\bH$, some error patterns tend to produce, on average, a larger (lower) amount of 
residual errors, after the first decoder iteration.
First of all, this statement is actually supported by empirical evidences: we have run numerical simulations on the same codes as 
those in Fig.s \ref{fig:ws_vs_dfr} and \ref{fig:subfigurew50_W40}, and have evaluated the number of residual errors after the first iteration. 
The results are shown in Fig. \ref{fig:num_errors_w50_w40}; as we can see, accordingly with Proposition \ref{pro:DFR_t_prime}, the trend of 
the \ac{DFR} and the one of $t'$ are the same for the analyzed codes.

We now derive a statistical model which approximates how the \ac{BF} decoder described in Algorithm \ref{alg:BF} evolves during the first iteration;
through this model we can predict the values of $t'$ and, thus, also justify the different trends of the \ac{DFR} observed in Fig.s 
\ref{fig:ws_vs_dfr} and \ref{fig:subfigurew50_W40}. 
We choose two distinct integers $i_0, i_1$ and consider the case of an error vector randomly drawn from the ensemble $\mathcal{E}(n,t,i_0,i_1)$, that is, 
we take columns $i_0$ and $i_1$ of $\bH$ as a reference, assuming that $e_{i_0}=e_{i_1}=1$. 
We also suppose that the columns $i_0$ and $i_1$ of $\bH$ overlap in $\gamma$ positions, and aim at expressing the average value of 
$t'$ as a function of the code parameters, the decoding threshold $b$ and the value of $\gamma$.
 
Let us first partition the sets of parity-check equations and variable nodes as follows.
\begin{Def}
Given an $r\times n$ parity-check matrix $\bH$, the set $\{0,1,\cdots,r-1\}$ can be partitioned into three subsets, defined as follows: 
\begin{enumerate}[label=\roman*)]
    \item $\mathcal{P}_0$: the set of parity-check equations that involve both bits $i_0$ and $i_1$, that is
    \begin{equation*}
        j\in\mathcal{P}_0 \text{\hspace{2mm}iff\hspace{2mm}}h_{j,i_0}=1\wedge h_{j,i_1}=1;
    \end{equation*}
    \item $\mathcal{P}_1$: the set of parity-check equations that involve either bit $i_0$ or bit $i_1$, that is
    \begin{equation*}
        j\in\mathcal{P}_1 \text{\hspace{2mm}iff\hspace{2mm}}(h_{j,i_0}=1\wedge h_{j,i_1}=0) \vee (h_{j,i_0}=0\wedge h_{j,i_1}=1);
    \end{equation*}
    \item $\mathcal{P}_2$: the set of parity-check equations that do not involve bits $i_0$ and $i_1$, that is
    \begin{equation*}
        j\in\mathcal{P}_2 \text{\hspace{2mm}iff\hspace{2mm}}h_{j,i_0}=0\wedge h_{j,i_1}=0.
    \end{equation*}
\end{enumerate}
\label{def:paritysets}
\end{Def}

\begin{Def}
 Let $\mathcal{V}_i$, with $i\in\{0,1,2\}$, be the set defined as
\begin{equation}
\mathcal{V}_i=\begin{cases}\left\{ \left. l\in [0;n-1]\setminus\{i_0,i_1\} \right| \exists j\in\mathcal{P}_i \text{\hspace{2mm}s.t.\hspace{2mm}}h_{j,l}=1\right\}&\text{if $i=0,1$}\\
\left\{ \left. l\in [0;n-1]\setminus\{i_0,i_1\}\right|  \not\exists j\in\mathcal{P}_0\cup\mathcal{P}_1 \text{\hspace{2mm}s.t.\hspace{2mm}}h_{j,l}=1\right\}&\text{if $i=2$}
\end{cases}.
\end{equation}
\label{def:vsets}
\end{Def}

The cardinality of each set $\mathcal{V}_i$ depends on the particular matrix $\bH$. However, when considering a regular code, we can derive some general properties, as stated in the following lemma.

\begin{Lem}
Given a $(v,w)$-regular code, the following bounds on the size of the sets $\mathcal V_i$, with $i\in \{0,1,2\}$, as defined in Definition \ref{def:vsets}, hold
\begin{equation}
\begin{cases}
    |\mathcal{V}_0|\leq \gamma(w-2),\\
    |\mathcal{V}_1|\leq (2v-2\gamma)(w-1),\\
    |\mathcal{V}_2|\geq n-2+\gamma w-2v(w-1).
    \end{cases}
    \label{eq:bounds}
\end{equation}
\end{Lem}
\begin{IEEEproof}
The first bound in \eqref{eq:bounds} follows from the fact that $|\mathcal{P}_0|=\gamma$. Any parity-check equation in $\mathcal{P}_0$ involves $w$ bits, including $i_0$ and $i_1$. So, all the parity-check equations in $\mathcal{P}_0$ involve, at most, $\gamma(w-2)$ bits other than $i_0$ and $i_1$. 
The second bound in \eqref{eq:bounds} can be derived with similar arguments, considering that  $|\mathcal{P}_1|=2(v-\gamma)$, and either $i_0$ or $i_1$ participates to the parity-check equations in $\mathcal{P}_1$.
The third bound is simply obtained by considering the remaining $r+\gamma-2v$ parity-check equations, when the other two bounds in \eqref{eq:bounds} hold with equality sign.
\end{IEEEproof}
From now on, in order to make our analysis as general as possible, i.e., independent of the particular $\bH$, we make the following assumption. 
\begin{Ass}
Let $\bH$ be the $r\times n$ parity-check matrix of a $(v,w)$-regular code.
We assume that each variable node other than $i_0$ and $i_1$ either participates in just one 
parity-check equation from $\mathcal{P}_0$ or $\mathcal{P}_1$ and $v-1$ equations in $\mathcal{P}_2$, or participates in $v$ parity-check 
equations in $\mathcal{P}_2$.
This means that
\begin{equation}
\begin{cases}
    |\mathcal{V}_0|= \gamma(w-2),\\
    |\mathcal{V}_1|= (2v-2\gamma)(w-1),\\
    |\mathcal{V}_2|= n-2+\gamma w-2v(w-1).
    \end{cases}
    \label{eq:boundseq}
\end{equation}
\label{ass:bounds}
\end{Ass}
The previous assumption is justified by the fact that, due to the sparsity of $\bH$, we have  $r\gg v$; it then follows that $|P_0|\ll |P_2|$ and $|P_1|\ll |P_2|$. 
Clearly, this assumption becomes more realistic as the matrix $\bH$ gets sparser.

We additionally define $t^{(i)}=\left|\phi(\be) \cap \mathcal{V}_i \right|$; clearly, we have $t^{(0)}+t^{(1)}+t^{(2)} = t-2$. 
The probability of having a specific configuration of $t^{(0)}$, $t^{(1)}$ and $t^{(2)}$ is equal to 
\begin{equation}
    P_{\{t^{(0)},t^{(1)},t^{(2)}\}} = \frac{\binom{|\mathcal{V}_0|}{t^{(0)}}\binom{|\mathcal{V}_1|)}{t^{(1)}}\binom{|\mathcal{V}_2|}{t^{(2)}}}{\binom{n-2}{t-2}}.
\end{equation}
In analogous way, we define $t^{(i)'}$ as the number of nodes that are in $\mathcal{V}^{(i)}$ and are simultaneously set in $\be+\be'_{[1]}$.
In other words, $t^{(i)'}$ corresponds to the number of errors that, after the first iteration, affect bits in $\mathcal{V}_i$.

The definitions of the sets $\mathcal{V}_i$ are useful to analyze how the value of $\gamma$ influences the decoder choices.
We focus on a generic $j$-th bit, with $j\neq i_0,i_1$, and consider the value of $\sigma_j$, as defined in Algorithm \ref{alg:BF}.
Because of Assumption \ref{ass:bounds}, we have that
\begin{enumerate}
 \item if $j\in\mathcal{V}_0$ (resp. $\mathcal{V}_1$), the $j$-th bit participates in one parity-check equation from $\mathcal{P}_0$ 
 (resp. $\mathcal{V}_1$) and $v-1$ parity-check equations in $\mathcal{P}_2$;
 \item if $j\in\mathcal{V}_2$, the $j$-th bit participates in $v$ parity-check equations in $\mathcal{P}_2$;
 \item if $j \in\{i_0, i_1\}$, then it participates in $\gamma$ parity-check equations from $\mathcal{P}_0$ and $v-\gamma$ parity-check equations from $\mathcal{P}_1$. 
\end{enumerate}
Let $p^{(i)}_{d,u}$, with $i=\{0,1,2\}$, be the probability that a parity-check equation involving the $j$-th bit (with $j\neq i_0,i_1$) and contained in $\mathcal{P}_i$ is unsatisfied, in the case of $e_j=d$, with $d\in\{0,1\}$; the value of such a probability is expressed by Lemma \ref{lem:probj}.
\begin{Lem}
Let us consider a $(v,w)$-regular code with blocklength $n$ and an error vector $\be$ with weight $t$. Then, the probabilities $p^{(i)}_{d,u}$, with $i\in\{0,1,2\}$ and $d\in\{0,1\}$, can be calculated as
\begin{equation}
\begin{gathered}
p^{(0)}_{0,u}=\sum_{\begin{smallmatrix}l=1\\
    \text{$l$ odd}\end{smallmatrix}}^{\min\{w-3,t-2\}}{\frac{\binom{w-3}{l}\binom{n-w}{t-l-2}}{\binom{n-3}{t-2}}},\\
    p^{(1)}_{0,u}=\sum_{\begin{smallmatrix}l=0\\
    \text{$l$ even}\end{smallmatrix}}^{\min\{w-2,t-2\}}{\frac{\binom{w-2}{l}\binom{n-w-1}{t-l-2}}{\binom{n-3}{t-2}}},\\
        p^{(0)}_{1,u}=\sum_{\begin{smallmatrix}l=0\\
    \text{$l$ even}\end{smallmatrix}}^{\min\{w-3,t-3\}}{\frac{\binom{w-3}{l}\binom{n-w}{t-l-3}}{\binom{n-3}{t-3}}},\\
    p^{(1)}_{1,u}=\sum_{\begin{smallmatrix}l=1\\
    \text{$l$ odd}\end{smallmatrix}}^{\min\{w-2,t-3\}}{\frac{\binom{w-2}{l}\binom{n-w-1}{t-l-3}}{\binom{n-3}{t-3}}},\\
    p^{(2)}_{0,u}=\sum_{\begin{smallmatrix}l=1\\
    \text{$l$ odd}\end{smallmatrix}}^{\min\{w-1,t-2\}}{\frac{\binom{w-1}{l}\binom{n-w-2}{t-l-2}}{\binom{n-3}{t-2}}},\\
    p^{(2)}_{1,u}=\sum_{\begin{smallmatrix}l=0\\
    \text{$l$ even}\end{smallmatrix}}^{\min\{w-1,t-3\}}{\frac{\binom{w-1}{l}\binom{n-w-2}{t-l-3}}{\binom{n-3}{t-3}}}.
    \end{gathered}
    \label{eq:probs}
\end{equation}
\label{lem:probj}
\end{Lem}
\begin{IEEEproof}
Let us first consider $i=0$ and $d=0$. Let us also consider the $j$-th bit, different from $i_0$ and $i_1$. Any parity check equation in $\mathcal{P}_0$ overlaps with the error vector in two positions, as it involves both bits $i_0$ and $i_1$; since we are looking at an error-free bit, then the parity-check equation will be unsatisfied only if the remaining $t-2$ errors intercept an odd number of ones, among the remaining $w-3$ ones. Simple combinatorial arguments lead to the first expression of \eqref{eq:probs}. All the other expressions can be derived with similar arguments.
\end{IEEEproof}
We also define $p^{(i)}_{\mathcal{E},u}$, with $i \in \{0,1\}$, as the probability that a parity-check equation involving a bit $\in\{i_0,i_1\}$, and contained in $\mathcal{P}_i$, is unsatisfied; the value of such a probability is derived in Lemma \ref{lem:LemmaE}.
\begin{Lem}
Let us consider a $(v,w)$-regular code with blocklength $n$ and an error vector $\be$ with weight $t$. Then, the probabilities $p^{(i)}_{\mathcal{E},u}$, with $i\in\{0,1\}$, can be calculated as 
\begin{equation}
    p^{(0)}_{\mathcal{E},u} = \sum_{\begin{smallmatrix}l=1\\\text{$l$ odd}
    \end{smallmatrix}}^{\min\{w-2,t-2\}}{\frac{\binom{w-2}{l}\binom{n-w}{t-2-l}}{\binom{n-2}{t-2}}},
\end{equation}
\begin{equation}
    p^{(1)}_{\mathcal{E},u} = \sum_{\begin{smallmatrix}l=0\\\text{$l$ even}
    \end{smallmatrix}}^{\min\{w-1,t-2\}}{\frac{\binom{w-1}{l}\binom{n-w-1}{t-2-l}}{\binom{n-2}{t-2}}}.
\end{equation}
\label{lem:LemmaE}
\end{Lem}
\begin{IEEEproof}
The proof can be carried on with the same arguments of the proof of Lemma \ref{lem:probj}.
\end{IEEEproof}
We now consider the following assumption.
\begin{Ass}
Let $\bH$ be the parity-check matrix of a $(v,w)$-regular code. We assume that the parity-check equations in which the $j$-th bit is involved are statistically independent; thus, $\sigma_j$, defined as in Algorithm \ref{alg:BF}, can be described in the first decoding iteration as the sum of independent Bernoulli random variables, each one having its own probability of being set, which corresponds either to $p^{(i)}_{d,u}$ or $p^{(d)}_{\mathcal{E},u}$, where $i\in\{0,1,2\}$ and $d\in\{0,1\}$. \label{ass:bernoulli_counter}
\end{Ass}

We now define $P^{(i)}_{d,\mathrm{flip}}$ as the probability that the decoder flips the $j$-th bit, in the case that $j\neq i_0,i_1$  and $j\in\mathcal{V}_i$, when $e_j=d$.
In analogous way, $P_{\mathcal{E},\mathrm{flip}}$ denotes the probability that, when $j\in\{i_0,i_1\}$, the decoder flips the $j$-th bit.
The above probabilities are computed in Lemmas \ref{pro:probsV}, \ref{the:thei1i0}, respectively.

\begin{Lem}
Let us consider a $(v,w)$-regular code with blocklength $n$ and an error vector $\be$ with weight $t$; let $b$ denote the decoding threshold employed in the first iteration. 
Then, under Assumptions \ref{ass:bounds} and \ref{ass:bernoulli_counter}, the probabilities $P^{(i)}_{d,\mathrm{flip}}$, with $i\in\{0,1,2\}$ and $d\in\{0,1\}$, can be computed as follows
\begin{equation}
    P^{(0)}_{d,\mathrm{flip}} = P\{\sigma_j^{(2)}=b-1|e_j=d\}p^{(0)}_{d,u}+\sum_{l=b}^{v-1}{P\{\sigma_j^{(2)}=l|e_j=d\}},
\end{equation}
\begin{equation}
    P^{(1)}_{d,\mathrm{flip}} = P\{\sigma_j^{(2)}=b-1|e_j=d\}p^{(1)}_{d,u}+\sum_{l=b}^{v-1}{P\{\sigma_j^{(2)}=l|e_j=d\}}.
\end{equation}
\begin{equation}
    P^{(2)}_{d,\mathrm{flip}} = \sum_{l=b}^{v}{\binom{v}{l}\left(p^{(2)}_{
    d,u}\right)^l\left(1-p^{(2)}_{d,u}\right)^{v-l}}.
\end{equation}
\label{pro:probsV}
\end{Lem}
\begin{IEEEproof}
When $j\in\mathcal{V}_0$, the $j$-th bit is involved in one parity-check equation in $\mathcal{P}_0$ and $v-1$ equations in $\mathcal{P}_2$.
The probability that the decoder in the first iteration flips the $j$-th bit can be computed as
\begin{equation}
    P^{(0)}_{d,\mathrm{flip}} = P\{\sigma_j^{(2)}=b-1|e_j=d\}p^{(0)}_{d,u}+P\{\sigma_j^{(2)}\geq b|e_j=d\}.
\end{equation}
In particular, we have
\begin{equation}
    P\{\sigma_j^{(2)}=z|e_j=d\} = \binom{v-1}{z}\left(p^{(2)}_{d,u}\right)^z\left(1-p^{(2)}_{d,u}\right)^{v-1-z},
\end{equation}
so that
\begin{equation}
    P^{(0)}_{d,\mathrm{flip}} = P\{\sigma_j^{(2)}=b-1|e_j=d\}p^{(0)}_{d,u}+\sum_{l=b}^{v-1}{P\{\sigma_j^{(2)}=l|e_j=d\}}.
\end{equation}
Similarly, if $j\in\mathcal{V}_1$, then it is involved in one parity-check equation in $\mathcal{P}_1$ and $v-1$ equations in $\mathcal{P}_2$; thus, we have
\begin{equation}
    P^{(1)}_{d,\mathrm{flip}} = P\{\sigma_j^{(2)}=b-1|e_j=d\}p^{(1)}_{d,u}+\sum_{l=b}^{v-1}{P\{\sigma_j^{(2)}=l|e_j=d\}}.
\end{equation}
Finally, if $j\in\mathcal{V}_2$, then it is involved in $v$ parity-check equations in $\mathcal{V}_2$; using a similar reasoning as in the previous cases, we can write
\begin{equation}
    P^{(2)}_{d,\mathrm{flip}} = \sum_{l=b}^{v}{\binom{v}{l}\left(p^{(2)}_{
    d,u}\right)^l\left(1-p^{(2)}_{d,u}\right)^{v-l}}.
\end{equation}
This proves the lemma. 
\end{IEEEproof}
\begin{Lem}Let us consider a $(v,w)$-regular code with blocklength $n$ and an error vector $\be$ with weight $t$; let $b$ denote the decoding threshold employed in the first iteration. 
Then, under Assumptions \ref{ass:bounds} and \ref{ass:bernoulli_counter}, the probability $P_{\mathcal{E},\mathrm{flip}}$ can be computed as follows
\begin{equation}
    P_{\mathcal{E},\mathrm{flip}} = \sum_{l^{(0)}=0}^{\gamma}\sum_{l^{(1)}=b-l^{(0)}}^{v-\gamma}{P\{\sigma_{\mathcal{E}}^{(0)}=l^{(0)}\}P\{\sigma_{\mathcal{E}}^{(1)}=l^{(1)}\}},
    \label{eq:flipi0i1}
\end{equation}
where
\begin{equation}
\begin{cases}
    P\{\sigma_{\mathcal{E}}^{(0)}=l\} = \binom{\gamma}{l}\left(p_{\mathcal{E},u}^{(0)}\right)^l\left(1-p_{\mathcal{E},u}^{(0)}\right)^{\gamma - l},\\
    P\{\sigma_{\mathcal{E}}^{(1)}=l\} = \binom{v-\gamma}{l}\left(p_{\mathcal{E},u}^{(1)}\right)^l\left(1-p_{\mathcal{E},u}^{(1)}\right)^{v-\gamma - l}.
    \end{cases}
    \label{eq:sysi0i1}
\end{equation}
\label{the:thei1i0}
\end{Lem}
\begin{IEEEproof}
Eq. \eqref{eq:sysi0i1} derives from the fact that bits $i_0$ and $i_1$ participate in $\gamma$ parity-check equations in $\mathcal{P}_0$; furthermore, both $i_0$ and $i_1$ participate in $v-\gamma$ equations in $\mathcal{P}_1$ each. Then, \eqref{eq:flipi0i1} expresses the probability that the number of unsatisfied parity-check equations for bit $i_0$ or $i_1$ is not smaller than the threshold $b$. 
\end{IEEEproof}

In order to estimate the average number of bits flipped after one iteration, we have to consider all the possible configurations of the error vector $\be$. As for the bits which are not in $\mathcal{E}$ the average value of $t^{(i)'}$ can be computed as
\begin{equation}
    E\left[ t^{(i)'}\right] = t^{(i)}\left(1-P^{(i)}_{1,\mathrm{flip}}\right)+\left(|\mathcal{V}_i|-t^{(i)}\right)P^{(i)}_{0,\mathrm{flip}},
    \label{eq:tzprimo}
\end{equation}
and the average number of errors in all bits $\mathcal{V}=\bigcup_{i=0}^{2}\mathcal{V}_i$ can be estimated as
\begin{equation}
    E\left[ t'_{\mathcal{V}}\right]=\sum_{t^{(0)}=0}^{\min\{t-2,|\mathcal{V}_0|\}}\sum_{t^{(1)}=0}^{\min\{t^{(1)}+t^{(2)},|\mathcal{V}_1|\}}{E_{\{t^{(0)},t^{(1)},t^{(2)}\}}\cdot P_{\{t^{(0)},t^{(1)},t^{(2)}\}}},
\end{equation}
where $t^{(2)}=t-2-(t^{(0)}+t^{(1)})$ and $E_{\{t^{(0)},t^{(1)},t^{(2)}\}}=E\left[t^{(0)'}\right] + E\left[t^{(1)'}\right] + E\left[t^{(2)'}\right]$. 

Similarly, the average number of residual errors due to the bits in $\mathcal{E}$ can be derived as
\begin{equation}
    E[t'_{\mathcal{E}}]=2(1-P_{\mathcal{E},\mathrm{flip}}).
\end{equation}

We can finally obtain the average value of $t'$ over all bits in $\{0,1,\ldots,n-1\}$ as
\begin{equation}
\label{eq:average_t_pr}
    E\left[ t'\right] = E[t'_{\mathcal{E}}] + E\left[t'_{\mathcal{V}}\right].
\end{equation}

A comparison between the simulated average values of $t'$ and the theoretical ones is shown in Table \ref{tab:comparison_eprime} for the two codes already considered in Figures \ref{fig:subfigurew50_W40} and \ref{fig:num_errors_w50_w40}, as a function of $\gamma$.
As we can see, this model allows for a close prediction of the average value of $t'$ starting from the number of overlapping ones $\gamma$.
This also allows an accurate modeling of the behaviour of the number of errors (increasing or decreasing) as a function of $\gamma$.

\begin{table}[ht]
\centering
\begin{tabular}{|c|cc|cc|}\hline
\multirow{2}{*}{$\gamma$} & \multicolumn{2}{c|}{Code (a)} &
      \multicolumn{2}{c|}{Code (b)} \\
& Simulated & Theoretical & Simulated & Theoretical \\\hline\hline
0 & 49.59 & 48.98 & 33.63 & 33.64\\ 
1 & 49.33 & 48.76 & 33.72 & 33.73\\ 
2 &   49.07 & 48.55 & 33.80& 33.80 \\ 
3 &  48.81 & 48.34& 33.86 & 33.86 \\
4& 48.55 & 48.15 & \textbf{-} & \textbf{-}\\\hline
\end{tabular}
\caption{Average values of $t'$, for different values of $\gamma$. Code (a) and Code (b) are the same as those considered in Figures \ref{fig:subfigurew50_W40} and \ref{fig:num_errors_w50_w40}.}\label{tab:comparison_eprime}
\end{table}

\section{Other sources of information leakage}\label{sec:NewSideChannel}
In this section we show some additional information leaks that might be exploited by an adversary to gather information about the structure of the secret $\bH$.

The results in the previous section show how, on average, the number of residual errors after the first iteration can be associated to the number of overlapping ones between columns of $\bH$.
Then, if the opponent has access to $\be'_{[1]}$ (i.e., to the positions that have been flipped in the first iteration), he can succeed in recovering the values of $\gamma_{i,j}$.
Indeed, once $\be'_{[1]}$ is known, the opponent can compute the number of residual errors for each query as $\be+\be'_{[1]}$.
Basically, this statistical attack can be modeled through Algorithm \ref{alg:generic_reaction}, by assuming that the oracle's answer $y^{(i)}$ is $t'$, that is, to the weight of $\be+\be'_{[1]}$.
The results in the previous section clearly show that this procedure allows for the cryptanalysis of the system.

We point out that, in a practical scenario, the locations of the bits that have been flipped by the decoder can be estimated through some power analysis attack, as in~\cite{Fabsic2016}.
This information might be masked through proper implementation strategies; for instance, random permutations might be applied to the order of processing bits in the decoder.
This solution, which was proposed by the authors of~\cite{Fabsic2016} as a countermeasure to the attack they introduced in the same paper, is however likely not to be strong enough for guaranteeing prevention of other kinds of information leaks.

For instance, let us suppose that the oracle's reply in Algorithm \ref{alg:generic_reaction} is the weight of $\be'_{[1]}$, i.e., to the number of flips performed in the first iteration. In a real case scenario, estimating this quantity might not be too hard.
Indeed, each flip requires the update of the error vector (one operation) and the update of the syndrome ($v$ operations).
Thus, we might expect that the duration of the first iteration, and/or its power consumption, linearly increases with the weight of $\be'_{[1]}$.
It can be shown that also this quantity depends on the number of intersections between columns.

Indeed, let us recall the notation adopted in the previous section, and define as $N_{\mathcal{V},\mathrm{flip}}^{(i)}$ the average number of flips performed among nodes in $\mathcal{V}_i$.
We can write
\begin{equation}
N_{\mathcal V,\mathrm{flip}}^{(i)} = t^{(i)'}P_{1,\mathrm{flip}}^{(i)}+(|\mathcal{V}_i|-t^{(i)'})P_{0,\mathrm{flip}}^{(i)},    
\end{equation}
so that the average number of flips in $\mathcal{V}$ is
\begin{equation}
    N_{\mathcal{V},\mathrm{flip}} =  \sum_{t^{(0)}=0}^{\min\{t-2,|\mathcal{V}_0|\}}\sum_{t^{(1)}=0}^{\min\{t^{(1)}+t^{(2)},|\mathcal{V}_1|\}}{P_{\{t^{(0)},t^{(1)},t^{(2)}\}}\left[N^{(0)}_{\mathcal{V},\mathrm{flip}}+N^{(1)}_{\mathcal{V},\mathrm{flip}}+N^{(2)}_{\mathcal{V},\mathrm{flip}}\right]}.
    \label{eq:probtiming}
\end{equation}
where $t^{(2)}=t-2-(t^{(0)}+t^{(1)})$.

The average number of flips for the bits in $\mathcal{E}$ is equal to $N_{\mathcal{E},\mathrm{flip}}=2P_{\mathcal{E},\mathrm{flip}}$.
So, combining the effect of the above equations, we have
\begin{equation}
    N_{\mathrm{flip}}=N_{\mathcal{E},\mathrm{flip}}+N_{\mathcal{V},\mathrm{flip}}.
\end{equation}
The probabilities in \eqref{eq:probtiming} depend on the value of $\gamma$; so, statistical attacks based on this quantity are expected to be successful.
We have verified this intuition by means of numerical simulations, and the results are shown at the end of this section.

Another quantity that might leak information about the secret key is represented by the evolution of the syndrome weight during iterations.
Authors in \cite{Eaton2018} have shown that the weight of the initial syndrome $\bs=\bH\be^T$ reveals information about the secret key; making a little step forward, we show that this consideration is indeed general and holds also for the first iteration.
We model this attack by assuming that an adversary runs Algorithm \ref{alg:generic_reaction} and the oracle replies with the syndrome weight after the first iteration, i.e. the weight of $\bs' = \bH (\be +\be'_{[1]})^T$. In general, we expect $t'\ll n$. On the other hand, large values of $t'$ are associated to large weights of $\bs'$ as well.
Since we have verified in the previous sections that error vectors drawn from ensembles $\mathcal{E}(n,t,i_0,i_1)$ are associated to different values of $t'$, it follows that also the syndrome weight depends on the number of intersections between columns of $\bH$. 

We have verified all these ideas by means of numerical simulations. In particular, we have considered \ac{QC} codes, described by $\bH$ in the form \eqref{eq:privateH}, in the case of $n_0=2$, $p=4801$ and $v=45$.
\begin{figure}[!tbh]
\centering
\subfigure[]{\includegraphics[keepaspectratio, width = 5.5cm]{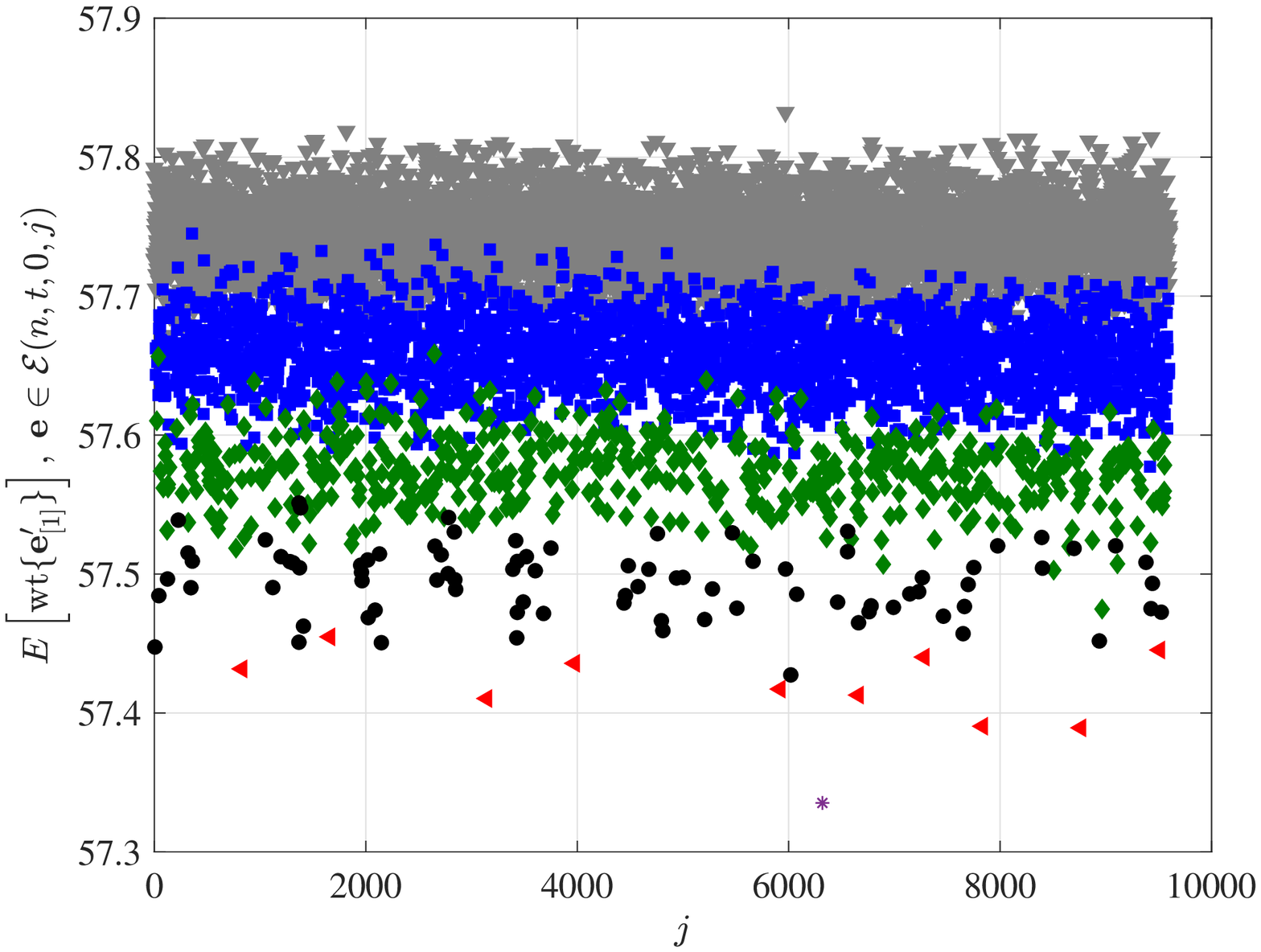}}\qquad
\subfigure[]{\includegraphics[keepaspectratio, width = 5.5cm]{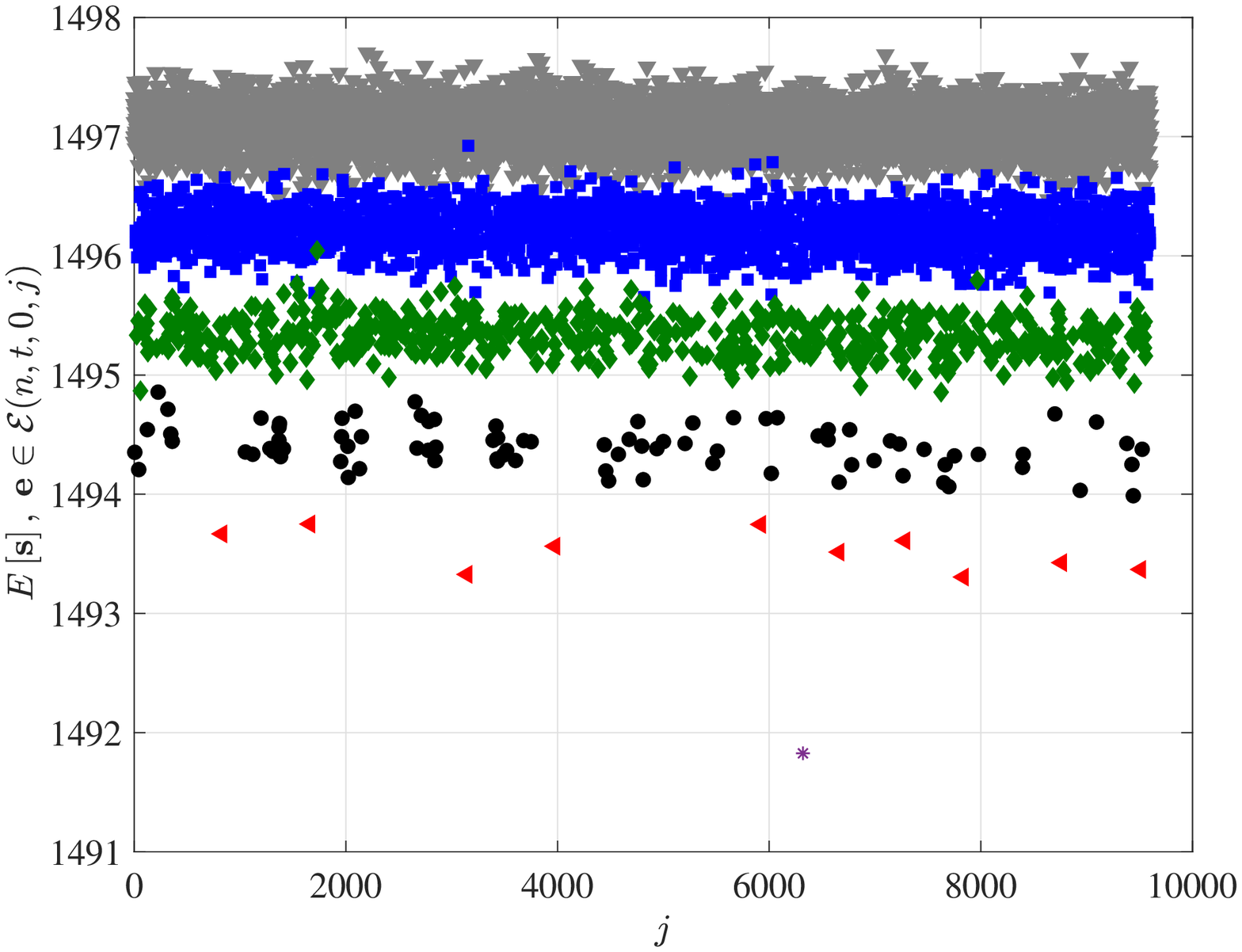}}\\
\caption{Simulation results for statistical attacks based on $t'$ and $\mathrm{wt}\{\bs'\}$.  
Figure (a) shows the distribution of the values of $t'$,  Figure (b) shows the weight of $s'$. 
The decoding threshold is $b=28$; results have been obtained through the simulation of $10^8$ decoding instances.
Grey, blue, green, black, red and violet markers are referred to couples of columns with number of intersections equal to $0,1,2,3,4,5$, respectively.
}\label{fig:new_attacks}
\end{figure}

We have chosen $t=84$ and $b=28$, and applied Algorithm~\ref{alg:generic_reaction} with $N=10^8$, considering that the oracle replies with: i) the average number of errors after the first iteration in Figure \ref{fig:new_attacks}(a), ii) the average weight of the syndrome vector after the first iteration in Figure \ref{fig:new_attacks}(b).
This empirical evidence confirms our conjectures, and proves that these data, when leaked, might lead to a complete cryptanalysis of the system.

\section{Conclusions\label{sec:Conclusion}}
In this paper we have provided a deep analysis of statistical attacks against \ac{LDPC} and \ac{MDPC} code-based cryptosystems.
We have considered a simple \ac{BF} decoder, and have shown that its probabilistic nature might yield a substantial information leakage.
We have shown that a general model for statistical attacks can be defined, and that many quantities, when observed by the opponent, might be exploited to recover information about the secret key.
Our analysis confirms that, in order to safely use long-lasting keys in McEliece cryptosystem variants based on sparse parity-check matrix codes, a constant time and power implementation is necessary, along with a negligible \ac{DFR}.
%
%
%
%
%

\bibliographystyle{splncs04}
\bibliography{References}

\begin{thebibliography}{10}
\providecommand{\url}[1]{\texttt{#1}}
\providecommand{\urlprefix}{URL }
\providecommand{\doi}[1]{https://doi.org/#1}

\bibitem{BIKE2017}
Aragon, N., Barreto, P.S.L.M., Bettaieb, S., Bidoux, L., Blazy, O., Deneuville,
  J.C., Gaborit, P., Gueron, S., G\"{u}neysu, T., Melchor, C.A., Misoczki, R.,
  Persichetti, E., Sendrier, N., Tillich, J.P., Zémor, G.: {BIKE}: Bit
  flipping key encapsulation (2017), \url{http://bikesuite.org/files/BIKE.pdf}

\bibitem{Baldi2008}
Baldi, M., Bodrato, M., Chiaraluce, F.: A new analysis of the {McEliece}
  cryptosystem based on {QC-LDPC} codes. In: Security and Cryptography for
  Networks, LNCS, vol.~5229, pp. 246--262. Springer Verlag (2008)

\bibitem{Baldi2007ISIT}
Baldi, M., Chiaraluce, F.: Cryptanalysis of a new instance of {McEliece}
  cryptosystem based on {QC-LDPC} codes. In: Proc. {IEEE} International
  Symposium on Information Theory (ISIT 2007). pp. 2591--2595. Nice, France
  (Jun 2007)

\bibitem{LEDAcrypt}
Baldi, M., Barenghi, A., Chiaraluce, F., Pelosi, G., Santini, P.: {LEDA}kem:
  {L}ow d{E}nsity co{D}e-b{A}sed key encapsulation mechanism (2017),
  \url{https://www.ledacrypt.org/}

\bibitem{Baldi2018}
Baldi, M., Barenghi, A., Chiaraluce, F., Pelosi, G., Santini, P.: {LEDA}kem: A
  post-quantum key encapsulation mechanism based on {QC-LDPC} codes. In: Lange,
  T., Steinwandt, R. (eds.) Post-Quantum Cryptography. pp. 3--24. Springer
  International Publishing, Cham (2018)

\bibitem{Becker2012}
Becker, A., Joux, A., May, A., Meurer, A.: Decoding random binary linear codes
  in $2^{n/20}$: How 1 + 1 = 0 improves information set decoding. In:
  Pointcheval, D., Johansson, T. (eds.) Advances in Cryptology - {EUROCRYPT}
  2012, LNCS, vol.~7237, pp. 520--536. Springer Verlag (2012)

\bibitem{Berlekamp1978}
Berlekamp, E., McEliece, R.J., van Tilborg, H.: On the inherent intractability
  of certain coding problems. {IEEE} Trans. Inf. Theory  \textbf{24}(3),
  384--386 (May 1978)

\bibitem{Bernstein2010}
Bernstein, D.J.: Grover vs. {M}c{E}liece. In: Sendrier, N. (ed.) Post-Quantum
  Cryptography. pp. 73--80. Springer Berlin Heidelberg, Berlin, Heidelberg
  (2010)

\bibitem{NISTreport2016}
Chen, L., Liu, Y.K., Jordan, S., Moody, D., Peralta, R., Perlner, R.,
  Smith-Tone, D.: Report on post-quantum cryptography. Tech. Rep. NISTIR 8105,
  National Institute of Standards and Technology (2016)

\bibitem{Eaton2018}
Eaton, E., Lequesne, M., Parent, A., Sendrier, N.: {QC-MDPC}: A timing attack
  and a {CCA}2 {KEM}. In: Lange, T., Steinwandt, R. (eds.) PQCrypto. pp.
  47--76. Springer International Publishing, Fort Lauderdale, FL, USA (Apr
  2018)

\bibitem{Fabsic2016}
Fab{\v{s}}i{\v{c}}, T., Gallo, O., Hromada, V.: Simple power analysis attack on
  the {QC-LDPC} {McEliece} cryptosystem. Tatra Mountains Math. Pub.
  \textbf{67}(1),  85--92 (Sep 2016)

\bibitem{Fabsic2017}
Fab{\v{s}}i{\v{c}}, T., Hromada, V., Stankovski, P., Zajac, P., Guo, Q.,
  Johansson, T.: A reaction attack on the {QC-LDPC} {McEliece} cryptosystem.
  In: Lange, T., Takagi, T. (eds.) Post-Quantum Cryptography: 8th International
  Workshop, PQCrypto 2017, pp. 51--68. Springer, Utrecht, The Netherlands (Jun
  2017)

\bibitem{Fabsic2018}
Fab{\v{s}}i{\v{c}}, T., Hromada, V., Zajac, P.: A reaction attack on {LEDA}pkc.
  IACR Cryptology ePrint Archive  \textbf{2018}, ~140 (2018)

\bibitem{Gallager1963}
Gallager, R.G.: Low-Density Parity-Check Codes. M.I.T. Press (1963)

\bibitem{Guo2016}
Guo, Q., Johansson, T., Stankovski, P.: A key recovery attack on {MDPC} with
  {CCA} security using decoding errors. In: Cheon, J.H., Takagi, T. (eds.)
  ASIACRYPT 2016, LNCS, vol. 10031, pp. 789--815. Springer Berlin Heidelberg
  (2016)

\bibitem{Kobara2001}
Kobara, K., Imai, H.: Semantically secure {McEliece} public-key cryptosystems
  --- conversions for {McEliece {PKC}}. LNCS  \textbf{1992},  19--35 (2001),
  \url{citeseer.ist.psu.edu/kobara01semantically.html}

\bibitem{Lee1988}
Lee, P., Brickell, E.: An observation on the security of {M}c{E}liece's
  public-key cryptosystem. In: Advances in Cryptology - EUROCRYPT 88, vol.~330,
  pp. 275--280. Springer Verlag (1988)

\bibitem{McEliece1978}
McEliece, R.J.: A public-key cryptosystem based on algebraic coding theory. DSN
  Progress Report pp. 114--116 (1978)

\bibitem{Misoczki2013}
Misoczki, R., Tillich, J.P., Sendrier, N., Barreto, P.S.L.M.:
  {MDPC}-{M}c{E}liece: New {M}c{E}liece variants from moderate density
  parity-check codes. In: 2013 IEEE International Symposium on Information
  Theory. pp. 2069--2073 (July 2013)

\bibitem{Niederreiter1986}
Niederreiter, H.: Knapsack-type cryptosystems and algebraic coding theory.
  Probl. Contr. and Inf. Theory  \textbf{15},  159--166 (1986)

\bibitem{Nilsson2018}
Nilsson, A., Johansson, T., Stankovski, P.: Error amplification in code-based
  cryptography. IACR Transactions on Cryptographic Hardware and Embedded
  Systems  \textbf{2019}(1),  238--258 (Nov 2018)

\bibitem{Paiva2018}
Paiva, T., Terada, R.: Improving the efficiency of a reaction attack on the
  {QC-MDPC} {M}c{E}liece. IEICE Transactions on Fundamentals of Electronics
  Communications and Computer Sciences  \textbf{E101.A},  1676--1686 (Oct 2018)

\bibitem{Prange1962}
Prange, E.: The use of information sets in decoding cyclic codes. IRE
  Transactions on Information Theory  \textbf{8}(5), ~5--9 (Sep 1962)

\bibitem{Santini2018ISIT}
Santini, P., Baldi, M., Cancellieri, G., Chiaraluce, F.: Hindering reaction
  attacks by using monomial codes in the {M}c{E}liece cryptosystem. pp.
  951--955 (Jun 2018)

\bibitem{Santini2018CANS}
Santini, P., Baldi, M., Chiaraluce, F.: Assessing and countering reaction
  attacks against post-quantum public-key cryptosystems based on {QC-LDPC}
  codes. In: Cryptology and Network Security - 17th International Conference,
  {CANS} 2018, Naples, Italy, September 30 - October 3, 2018, Proceedings. pp.
  323--343 (2018)

\bibitem{Stern1989}
Stern, J.: A method for finding codewords of small weight. In: Cohen, G.,
  Wolfmann, J. (eds.) Coding Theory and Applications, LNCS, vol.~388, pp.
  106--113. Springer Verlag (1989)

\bibitem{Tillich2018}
Tillich, J.P.: The decoding failure probability of {MDPC} codes. 2018 IEEE
  International Symposium on Information Theory (ISIT) pp. 941--945 (Jun 2018)

\end{thebibliography}

\end{document}